# Polarimetric and Photometric Observations of NEAs; (422699) 2000 PD3 and (3200) Phaethon with the 1.6m Pirka Telescope

***Authors***

Ryo Okazaki* **, Tomohiko Sekiguchi*

Hokkaido University of Education, 9 Hokumon, Asahikawa 070-0825, Japan*

Asahikawa MEISEI High School, 14 Midorimachi, Asahikawa 070-0832, Japan**

Masateru Ishiguro

Department of Physics and Astronomy, Seoul National University, Gwanak, Seoul 08826, South Korea

Hiroyuki Naito

Nayoro Observatory, 157-1 Nisshin, Nayoro, Hokkaido 096-0066, Japan

Seitaro Urakawa

Bisei Spaceguard Center, Japan Spaceguard Association, 1716-3 Okura, Bisei-cho, Ibara, Okayama 714-1411, Japan

Masataka Imai

Artificial Intelligence Research Center, National Institute of Advanced Industrial Science and Technology, Aomi, Koto-ku, Tokyo 136-0064, Japan

Tatsuharu Ono

Department of Cosmosceinces, Graduate School of Science, Hokkaido University, Kita-ku, Sapporo, Hokkaido 060-0810, Japan

Brian D. Warner

Center for Solar System Studies, Palmer Divide Station, 446 Sycamore Ave, Eaton, CO 80615, USA

Makoto Watanabe

Department of Applied Physics, Okayama University of Science, 1-1 Ridai-cho, Kita-ku, Okayama, Okayama 700-0005, Japan

***Abstract***

We report on optical polarimetric observations of two Apollo type near-Earth asteroids, (422699) 2000 PD3 and (3200) Phaethon, and BVRI photometric observations of 2000 PD3 using the 1.6m Pirka telescope in 2017. We derived the geometric albedo of $p_{\mathrm{v}} = 0.22 \pm 0.06$ and the color indices ($B$-$V$ = 0.282 ± 0.072, $V$-$R$ = 0.198 ± 0.035 and $V$-$I$ = 0.203 ± 0.022) for 2000 PD3 which are consistent with those of S-type asteroids (including Q-types). The effective diameter of 2000 PD3 was derived as $0.69 \pm 0.15$ km using our derived geometric albedo. We found that our polarimetric data of Phaethon in 2017 is deviated from the polarimetric profile taken at different epoch of 2016 using the identical instrument setting (Ito et al., 2018). This result suggests that Phaethon would have a regional heterogeneity in grain size and/or albedo on its surface.

## 1. Introduction

Polarimetric observing technique has been applied to study physical properties on asteroid surfaces because it is a powerful method for investigating the light scattering properties such as the albedo and the regolith particle size. The liner polarization degree $P_r$ is defined as

$$P_r = \frac{I_\perp - I_{/\!/}}{I_\perp + I_{/\!/}}, \tag{1}$$

where $I_\perp$ and $I_{/\!/}$ are the perpendicular and parallel components of the intensities of light beam with respect to the scattering plane (the plane where the Sun, asteroid and observers exist), respectively. It has been known that $P_r$ exhibits a strong dependence on the phase angle $\alpha$ (Sun-Target-Observer's angle) (Dollus et al., 1989). Such relation is characterized by the presence of a negative branch and a positive branch (Geake and Dollfus, 1986). The geometric albedo of asteroids in the V-band, $p_v$ can be derived empirically (but reliably) from polarimetric slopes around the inversion angle (Zellner and Gradie 1976) although that is often obtained by the combination of thermal infrared observations and optical observations (e.g. Sekiguchi et al., 2018). The albedo value is of importance for estimating diameters and classifying taxonomic types of asteroids. Besides, the maximum values of the polarization degree, $P_{max}$ have a moderate correlation with the surface grain sizes as well as $p_v$ (so-called Umov law, Bowell et al., 1972; Dollfus 1998; Shkuratov and Opanasenko 1992). Despite such importance of the polarimetry, there are a limited number of polarimetric studies on asteroids at large phase angles ($\alpha \geq 30°$) because of limited observing opportunities. Only near-Earth asteroids allow us to observe at larger phase angles because they have the perihelia inside the orbits of the Earth.

We report our new optical polarimetric observations of two Apollo type asteroids, (422699) 2000 PD3 and (3200) Phaethon and photometric observations of 2000 PD3 at relatively large phase angles in order to understand the scattering properties of these NEAs. We summarized their orbital parameters in Table 1. In autumn of 2017, 2000 PD3 was observable over the wide range ($\alpha$ = 20°-120°). Phaethon gains attention to planetary scientists in that it is the parent body of the Geminid meteor stream. This asteroid is classified as F-type (Tholen 1984) or B-type (Bus and Binzel 2002). In December 2017, there was a unique opportunity to obtain polarimetric data at extremely large phase angle ($\alpha$ = 30°-130°). The major purpose of this study was a polarimetric comparison of an S-type asteroid and a C/B-type asteroid. Within the period, we got one-night polarimetric data to confirm consistency with polarimetric data taken with the identical instrument used one year prior to our observations (Ito et al., 2018).

The observations and the data reduction are described in Section 2, the results in Section 3 and the discussion in Section 4. In this paper, we mostly focus on the derivation of albedo, size and taxonomic type of 2000 PD3 because we obtained a substantial data set. However, we would emphasis that our follow-up polarimetric observation of Phaethon using the instrumental setting identical to Ito et al. (2018) is also important, because the regional heterogeneity on the asteroid has just been involved in an argument (Zheltobryukhov et al. 2018; Borisov et al. 2018; Shinnaka et al. 2018; Devogèle et al. 2018).

Table 1. Orbital parameters of two target NEAs

| Orbital Elements | 2000 PD3 | Phaethon |
|---|---|---|
| $e$: eccentricity | 0.593 | 0.890 |
| $a$: semimajor axis (au) | 1.997 | 1.271 |
| $q$: perihelion distance (au) | 0.812 | 0.140 |
| $Q$: aphelion distance (au) | 3.182 | 2.403 |
| $i$: inclination (deg.) | 7.689 | 22.221 |
| P: orbital period (year) | 2.82 | 1.43 |
| perihelion passage (UT) | 2017-Oct-19 | 2018-Jan-25 |
| Earth approach (UT) | 2017-Sep-3 | 2017-Dec-16 |
| closest approach distance (au) | 0.110 | 0.069 |

From JPL Small-Body Database Browser of 2019-Apr.-27.0[1]

[1] https://ssd.jpl.nasa.gov/sbdb.cgi

## 2. Observations and Data Reduction

Our targets were selected for comparison of a typical S-type redder asteroid (2000 PD3) and a typical C/B-type bluer asteroid (Phaethon) in their polarimetric aspects.

Polarimetric and photometric observations were performed in 2017 using the 1.6 m Pirka telescope and Multi-Spectral Imager (MSI) (Watanabe et al., 2012) mounted on the Cassegrain focus of the Pirka telescope at the Nayoro Observatory of Hokkaido University, Japan. This instrument enables the acquisition of images that cover a 3.3′×3.3′ field of view (FOV) with 0.39′′ pixel resolution. The Johnson Cousins filters of B, V, $R_c$ and $I_c$ are adopted. Hereafter, we describe the details about photometry and polarimetry separately.

### *2.1 Polarimetry*

Polarimetric observations of two near-Earth asteroids, 2000 PD3 and Phaethon were conducted from UT 2017 August 15 to UT 2017 December 16. The imaging polarimetric mode of MSI which implements a rotatable half wave plate as the polarimetric modulator and the Wollaston prism as the beam splitter, generates simultaneous ordinary and extraordinary images with mutually perpendicular polarizations at four position angles ($\theta$ = 0°, 45°, 22.5°, and 67.5°) (Watanabe et al., 2012). The observing geometry is summarized in Table 2. The exposure lengths for the same objects vary a lot even for obtaining more S/N. Polarimetric data and the standard calibration frames (e.g., biases and dome flat fields) were acquired for each observation.

Since the detailed methods for data analysis and derivation of errors are described in Ishiguro et al. (2017), Kuroda et al. (2017) and Ito et al. (2018), we explain the outline. The basic reduction such as the flat-fielding were processed by the MSI data reduction package (MSIRED [Watanabe et al., 2012]) using the IRAF. In order to measure the intensities of ordinary rays and extraordinary rays, the aperture photometry method was applied for all images of each asteroid and polarimetric standard star (HD212311). The Stokes parameters (*I*, *Q* and *U*) are deduced from these intensities at a set of half-wave plate position angles in degree, which are given as

$$\frac{Q}{I} = \left(\sqrt{\frac{I_{e0}/I_{o0}}{I_{e45}/I_{o45}}}-1\right) / \left(\sqrt{\frac{I_{e0}/I_{o0}}{I_{e45}/I_{o45}}}+1\right), \tag{2}$$

and

$$\frac{U}{I} = \left(\sqrt{\frac{I_{e22.5}/I_{o22.5}}{I_{e67.5}/I_{o67.5}}}-1\right) / \left(\sqrt{\frac{I_{e22.5}/I_{o22.5}}{I_{e67.5}/I_{o67.5}}}+1\right), \tag{3}$$

where the subscript characters ‘o’ and ‘e’ denote the ordinary and extraordinary components while subscript numbers stand for the angles (in degree) of half wave plate. The liner polarization degree ($P$) and the position angle of polarization ($\theta_p$) are calculated from

$$P = \sqrt{\left(\frac{Q}{I}\right)^2 + \left(\frac{U}{I}\right)^2}, \tag{4}$$

and

$$\theta_p = \frac{1}{2}\tan^{-1}\left(\frac{U}{Q}\right), \quad (5)$$

After correcting for instrumental effects which include corrections for the polarization efficiency, the instrumental polarization and position angle offset (see, Appendix), the liner polarization degree ($P_r$) and the position angle of polarization ($\theta_r$) with respect to the scattering plane (Zellner and Gradie, 1976) are derived by

$$P_r = P\cos(2\theta_r), \quad (6)$$

and

$$\theta_r = \theta_p - (\Phi \pm 90°), \quad (7)$$

where $\Phi$ is the position angle of the scattering plane on the sky.

Table 2. Observing geometry and results of our polarimetric observations

| Object | Date (UT) | Filter | $t_{exp}$[a] (s) | $N$[b] | $R_h$[c] (au) | $\Delta$[d] (au) | $\alpha$[e] (deg.) | $\Phi$[f] (deg.) | Airmass | $P$[g] $\pm$ $\varepsilon P$[h] (%) | $\theta_p$[i] $\pm$ $\varepsilon\theta_p$[j] (deg.) | $P_r$[k] (%) | $\theta_r$[l] (deg.) |
|---|---|---|---|---|---|---|---|---|---|---|---|---|---|
| 2000 PD3 | 2017-Aug-15 13:53 - 14:50 | $R_c$ | 60 | 40 | 1.200 | 0.211 | 25.5 | 206.7 | 1.250 - 1.193 | 1.37 ± 0.45 | 41.70 ± 9.93 | 1.18 | 15.03 |
| | 2017-Aug-15 14:50 - 14:55 | V | 60 | 4 | 1.200 | 0.211 | 25.5 | 206.7 | 1.176 | 2.69 ± 2.01 | 24.23 ± 32.12 | 2.68 | 2.44 |
| | 2017-Aug-29 14:15 - 14:44 | $R_c$ | 60 | 28 | 1.077 | 0.120 | 53.5 | 185.3 | 1.008 - 1.004 | 3.66 ± 0.36 | 15.48 ± 2.85 | 3.43 | 10.15 |
| | 2017-Aug-29 14:59 - 15:03 | V | 60 | 4 | 1.077 | 0.120 | 53.5 | 185.3 | 1.005 | 3.11 ± 1.43 | 10.28 ± 14.91 | 3.06 | 4.95 |
| | 2017-Sep-4 10:51 - 12:18 | $R_c$ | 90 | 40 | 1.028 | 0.110 | 76.9 | 190.8 | 1.328 - 1.241 | 6.68 ± 0.56 | 18.10 ± 2.39 | 6.46 | 7.30 |
| | 2017-Sep-4 12:40 - 12:59 | V | 90 | 12 | 1.028 | 0.110 | 76.9 | 190.8 | 1.225 | 7.18 ± 1.07 | 22.28 ± 4.33 | 6.61 | 11.47 |
| | 2017-Sep-5 10:03 - 10:53 | $R_c$ | 90 | 32 | 1.020 | 0.111 | 80.8 | 197.1 | 1.420 - 1.371 | 6.26 ± 0.73 | 28.55 ± 3.36 | 5.77 | 11.43 |
| | 2017-Sep-5 11:13 - 11:20 | V | 90 | 4 | 1.020 | 0.111 | 80.8 | 197.1 | 1.352 | 8.29 ± 2.85 | 18.18 ± 10.49 | 8.28 | 1.06 |
| | 2017-Sep-6 11:10 - 11:48 | $R_c$ | 90 | 16 | 1.012 | 0.112 | 84.9 | 214.1 | 1.419 - 1.397 | 8.33 ± 1.60 | -36.42 ± 5.61 | 8.31 | -2.30 |
| Phaethon | 2017-Dec-14 09:19 - 09:23 | $R_c$ | 10 | 12 | 1.053 | 0.084 | 33.4 | 112.3 | 1.137 - 1.130 | 2.98 ± 0.22 | 30.61 ± 2.10 | 2.86 | 8.31 |
| | 2017-Dec-14 09:21 - 09:27 | V | 10 | 16 | 1.053 | 0.084 | 33.4 | 112.3 | 1.134 - 1.122 | 1.71 ± 0.22 | 33.74 ± 3.67 | 1.58 | 11.44 |

| | | | | | | | | | | | | |
|---|---|---|---|---|---|---|---|---|---|---|---|---|
| 2017-Dec-16 10:54 - 10:59 | $R_c$ | 20 | 12 | 1.021 | 0.069 | 57.9 | 79.0 | 1.051 | 14.53 ± 0.11 | 8.35 ± 0.22 | 14.47 | -2.68 |
| 2017-Dec-16 11:01 - 11:05 | V | 20 | 12 | 1.021 | 0.069 | 57.9 | 79.0 | 1.058 | 15.06 ± 0.13 | 8.70 ± 0.26 | 15.02 | -2.28 |

[a] Exposure time per frame in seconds.

[b] Number of exposures.

[c] Heliocentric distance in au.

[d] Geocentric distance in au.

[e] Solar phase angle in degrees.

[f] Position angle of the scattering plane in degrees.

[g] Polarization degree in percent.

[h] Error of polarization degree in percent.

[i] Position angle of strongest electric vector in degrees.

[j] Error of position angle of strongest electric vector in degrees.

[k] Polarization degree with respect to the scattering plane in percent.

[l] Position angle with respect to the scattering plane in degrees.

*2.2 Photometry*

BVRI photometric observations of 2000 PD3 were carried out on two photometric nights, UT 2017-Aug-29 and UT 2017-Sep-4. Photometric data of the target asteroid 2000 PD3 and the Landolt standard stars, SA20 245 and SA20 139 (Landolt 2013), were acquired at airmass as close as possible to cancel out the wavelength-dependent atmospheric extinction. Because the difference in airmass was so small (0.01-0.04), photometric errors were smaller than about 0.01 magnitude (0.003-0.012 mag) under the typical sky condition at the observatory (i.e., the extinction coefficient of ~0.3 was assumed). Bias frames and dome flat field frames were taken every night. Table 3 presents the observing geometry of the object and standard stars. The basic reduction and aperture photometry were executed in the same way as polarimetry (Subsection 2.1).

Table 3. Observing geometry of BVRI photometry

| Object | Date (UT) | Filter | $t_{exp}$[a] (s) | $N$[b] | $r$[c] (au) | $\Delta$[d] (au) | $\alpha$[e] (deg.) | Airmass |
|---|---|---|---|---|---|---|---|---|
| 2000 PD3 | 2017-Aug-29 24:44 | V | 60 | 1 | 1.200 | 0.211 | 25.5 | 1.018 |
| | 2017-Aug-29 24:45 | $R_c$ | 60 | 1 | 1.200 | 0.211 | 25.5 | 1.019 |
| | 2017-Aug-29 24:48 | $I_c$ | 60 | 1 | 1.200 | 0.211 | 25.5 | 1.020 |
| | 2017-Aug-29 24:49 | B | 90 | 1 | 1.200 | 0.211 | 25.5 | 1.021 |
| SA 20 245 | 2017-Aug-29 24:54 | V | 0.5 | 1 | | | | 1.011 |
| | 2017-Aug-29 24:55 | $R_c$ | 0.5 | 1 | | | | 1.011 |
| | 2017-Aug-29 24:56 | $I_c$ | 0.5 | 1 | | | | 1.011 |
| | 2017-Aug-29 24:57 | B | 1.0 | 1 | | | | 1.010 |
| 2000 PD3 | 2017-Sep-4 22:06 | $R_c$ | 60 | 1 | 1.028 | 0.110 | 76.9 | 1.209 |
| | 2017-Sep-4 22:08 | V | 60 | 1 | 1.028 | 0.110 | 76.9 | 1.208 |
| | 2017-Sep-4 22:10 | $I_c$ | 60 | 1 | 1.028 | 0.110 | 76.9 | 1.208 |
| | 2017-Sep-4 22:17 | B | 90 | 1 | 1.028 | 0.110 | 76.9 | 1.207 |
| SA 20 139 | 2017-Sep-4 22:23 | $R_c$ | 10 | 1 | | | | 1.164 |
| | 2017-Sep-4 22:26 | V | 10 | 1 | | | | 1.159 |
| | 2017-Sep-4 22:27 | $I_c$ | 10 | 1 | | | | 1.157 |
| | 2017-Sep-4 22:28 | B | 10 | 1 | | | | 1.155 |

[a] Exposure time per frame in seconds.

[b] Number of exposures.

[c] Heliocentric distance in au.

[d] Geocentric distance in au.

[e] Solar phase angle in degrees.

## 3. Results

### *3.1 Linear Polarization Degree*

We summarize the results of our polarimetric observation in Table 2. The error values of $P_{\mathrm{r}}$ and $\theta_{\mathrm{r}}$ are same as those of $P$ and $\theta_{\mathrm{p}}$, respectively. We obtained the polarimetric results at five phase angles ($\alpha$ = 25.5°-84.9°) in the $\mathrm{R_c}$ band and four phase angles ($\alpha$ = 25.5°-80.8°) in the V band for 2000 PD3. In Fig.1, the $P_{\mathrm{r}}$ values of 2000 PD3 in both bands tend to increase with the phase angles and display those of S-types. The measurements of Phaethon were carried out for two nights at $\alpha$ = 33.4°-57.9° and showed the high polarization (e.g., $P_{\mathrm{r}}$ = 15.02 ± 0.13% at $\alpha$ = 57.9°). Phaethon has higher polarizations than those of S-types and C-types (see, Fig.1)

All observations were made in the same setting as Ishiguro et al. (2017) and Ito et al. (2018). Note here that our polarimetric data are reliable within the derived errors despite the variable sky conditions because the beam splitter type (i.e. Wollaston prism) polarimetric data can essentially cancel the time-dependent weather factors.

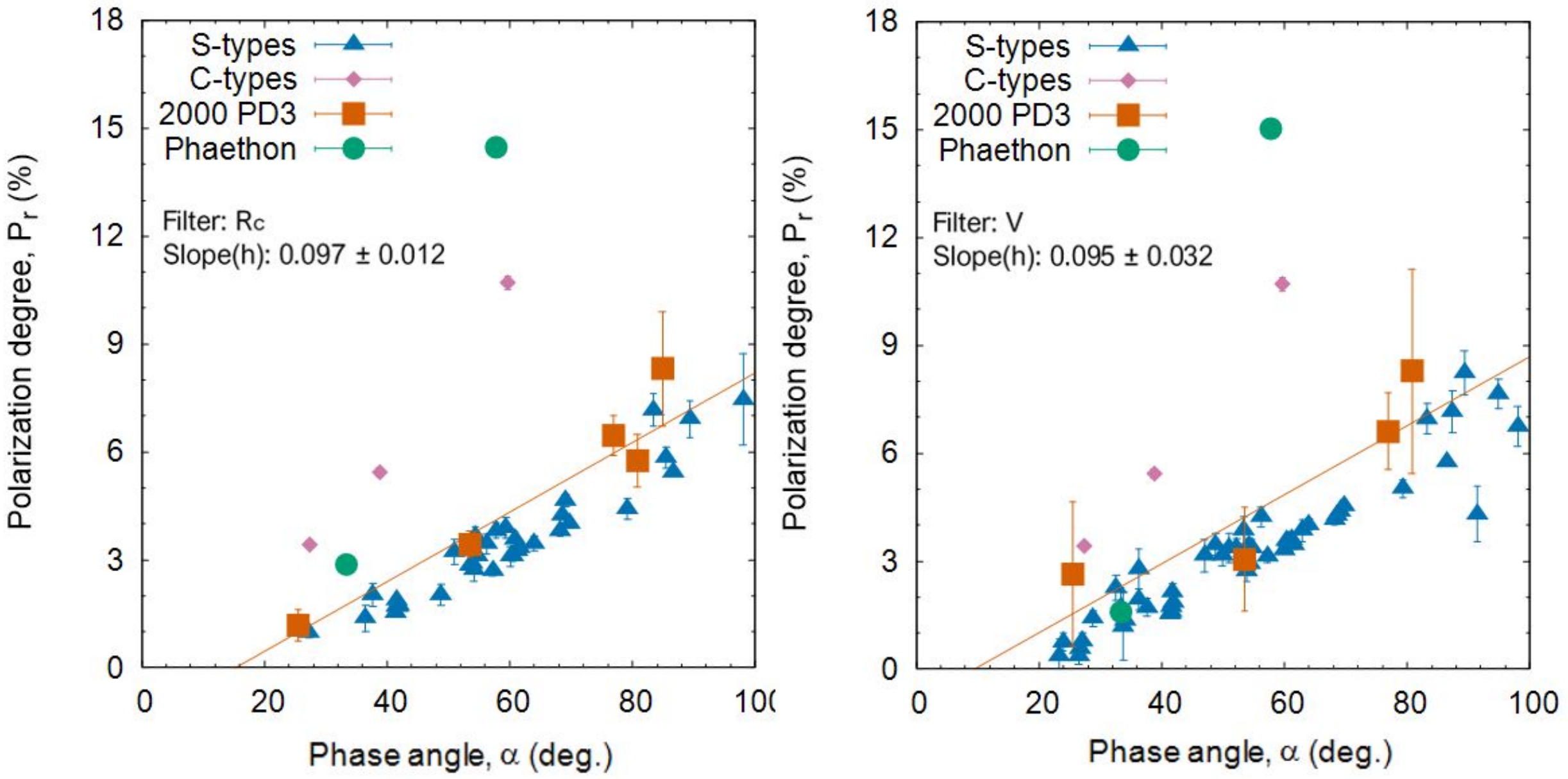

Figure 1. Phase angle $\alpha$ - polarization degree $P_r$ relation of 2000 PD3, Phaethon and the other asteroids. The figure to the left is the data in $R_c$-band and the right is the data in V-band. The triangles denote S-type asteroids:1685 Toro (Kiselev et al., 1990), 1036 Ganymed (Kiselev et al., 1994), 25143 Itokawa (Cellino et al., 2005) and Q-type asteroid: 1566 Icarus (Ishiguro et al., 2017). The rhombus symbols show C-type asteroids: 2100 Ra-Shalom (Kiselev et al., 1999) and 1580 Betulia (Tedesco et al., 1978). The squares and the circles represent data for our observations of 2000 PD3 and Phaethon. These solid lines show weighted liner fits of the obtained data of 2000 PD3.

### *3.2 Geometric Albedo of 2000 PD3*

The geometric albedo $p_v$ of 2000 PD3 was measured from its thermal infrared to be 0.22 ± 0.18 by NEOWISE (Nugent et al., 2016) using the NEATM (Near Earth Asteroid Thermal Model; Harris 1998). On the other side, it can be also derived from polarimetric measurement using the empirical relationship so-called "the slope-albedo law", which is based on scattering properties of asteroids (Zellner and Gradie 1976; Lupishko and Mohamed 1996; Celiino et al., 1999) and given by

$$\log_{10}(p_v) = C_1 \log_{10}(h_{slp}) + C_2, \tag{8}$$

where $C_1$ and $C_2$ are constants. $C_1 = -0.780 \pm 0.037$ and $C_2 = -1.469 \pm 0.036$ (when $p_v \geq 0.08$) were derived by Cellino et al. (2015). $h_{slp}$ is the polarimetric slope near the inversion angle.

The polarimetric slope $h_{slp} = 0.095 \pm 0.032$ (Fig. 1) was obtained from our V-band data of 2000 PD3 and $p_v = 0.22 \pm 0.06$ was calculated using the formula (8). We adopted the linear fitting which should be reasonable in the large phase angle range (e.g., Cellino et al. 2005; Ishiguro et al., 2017). The geometric albedo in the $R_c$-band: $p_R = 0.21 \pm 0.03$ was derived from our $R_c$-band data with the same manner as V-band. It is converted into $p_v = 0.24 \pm 0.04$ using the *V-R* color index ($0.164 \pm 0.019$) on August 29, 2017. Table 4 lists our results and the polarimetric mean value for each taxonomy. Our obtained values match those of S-types (0.207 ± 0.089; Usui et al., 2013) and agree well with a recent polarimetric study (Belskaya et al., 2017) which have presented mean polarization parameters of S-type and Q-type asteroids. The inversion angles (~16° in the $R_c$-band and ~10° in the V-band) are slightly smaller than S-type asteroids (a typical value is 20.7°) in Fig 1.

Table 4. Comparison of the geometric albedos

| Name | $p_v$ | Note |
|---|---|---|
| 2000 PD3 | 0.22 ± 0.06 | This study |
| 2000 PD3 | 0.21 ± 0.03 | This study ($p_R$) |
| 2000 PD3 | 0.24 ± 0.04 | This study (Converted from its $p_R$) |
| 2000 PD3 | 0.22 ± 0.18 | Thermal infrared technique (Nugent et al., 2016) |
| S-types | 0.207 ± 0.089 | Thermal infrared mean value (Usui et al., 2013) |
| S-types | 0.235 ± 0.046 | Polarimetric mean value (Belskaya et al., 2017) |
| Q-types | 0.29 ± 0.08 | Polarimetric mean value (Belskaya et al., 2017) |
| C-types | 0.065 ± 0.015 | Polarimetric mean value (Belskaya et al., 2017) |
| V-types | 0.34 | Polarimetric mean value (Belskaya et al., 2017) |

*3.3 Diameter of 2000 PD3*

In order to estimate the diameter of 2000 PD3, its absolute magnitude and geometric albedo are needed. The absolute magnitude $H$ of asteroids is the V-magnitude if they would be placed at 1 au from the geocenter and the Sun, and at zero phase angle. The absolute magnitude is expressed as a so-called the H-G system (Bowell et al. 1989) and given by

$$H = H(\alpha) + 2.5\log_{10}[(1 - G)\Phi_1(\alpha) + G\Phi_2(\alpha)], \quad (9)$$

where $H(\alpha)$ is the reduced magnitude at the phase angle α, and the slope parameter $G$. $\Phi_1(\alpha)$ and $\Phi_2(\alpha)$ are functions that describe the scattering off the surface (Dymock, 2007). We calculated $\Phi_1(\alpha) = 0.056$ and $\Phi_2(\alpha) = 0.244$. We applied the empirical value $G = 0.15$ (e.g., Bowell et al. 1989), which is typical to S-type asteroids. The reduced visual magnitude is expressed as

$$H(\alpha) = V_{\rm obs}(\alpha) - 5\log_{10}[r\Delta], \quad (10)$$

where $r$ and Δ are the heliocentric and geocentric distances of the object, respectively. For estimating the $H$ value of this object, we just used the observational result on September 4 which corresponds to the rotation phase = 0.76 (see, Fig. 2). The reduced magnitude of this object at this point seems an average of the variation magnitude of this object in Fig. 2. Thus, we calculated $H = 18.11 \pm 0.050$ (magnitude) from the apparent magnitude $V_{\mathrm{obs}}(76.9) = 16.06 \pm 0.050$ at the observing run on September 4, 2017.

The absolute magnitude *H*, geometric albedo $p_v$ and diameter *D* (km) for asteroids are related by the following relation (e.g., Fowler and Chillemi, 1992).

$$D = \frac{1329 \times 10^{-0.2H}}{\sqrt{p_{\mathrm{v}}}} \tag{11}$$

The effective diameter of 2000 PD3 was derived as $0.69 \pm 0.15$ km using the polarimetric geometric albedo (Subsection 3.2), which is in a good agreement with the NEOWISE estimating as 0.62 ± 0.24 km (Nugent et al., 2016). The surface regolith of such small size asteroids may consist of large size particles. The scenario support that the above-mentioned small inversion angles are associated with large regolith, as we have seen on sub-km S-type asteroid Itokawa (Yano et al. 2006).

### *3.4 BVRI colors of 2000 PD3*

Table 5 presents colors of 2000 PD3 in terms of color indices (*B-V*, *V-R*, *V-I*) that are representing coarse measures of the surface taxonomy of asteroids. The color values are corrected for the influence of the solar color (*B-V* = 0.642, *V-R* = 0.354, *V-I* = 0.688; Holmberg et al. 2006). Fig. 3 shows *V-R* vs. *B-V* relation (the left panel) and *V-R* vs. *V-I* relation (the right panel) of 2000 PD3, respectively, with those of C-type, D-type, S-type, V-type and X-type asteroids. The three broadband colors of 2000 PD3 are plotted together within the those of S-type asteroids. Although nightly variation was noticed probably because of the color heterogeneity on the asteroid, we can safely insist that 2000 PD3 exhibits the color indices consistent with S-type asteroids.

Table 5.  Color indices of 2000 PD3

| Date | *B-V* (mag.) | *V-R* (mag.) | *V-I* (mag.) |
|---|---|---|---|
| 2017-Aug-29 | 0.180 ± 0.020 | 0.164 ± 0.019 | 0.168 ± 0.011 |
| 2017-Sep-4 | 0.282 ± 0.072 | 0.198 ± 0.035 | 0.203 ± 0.022 |

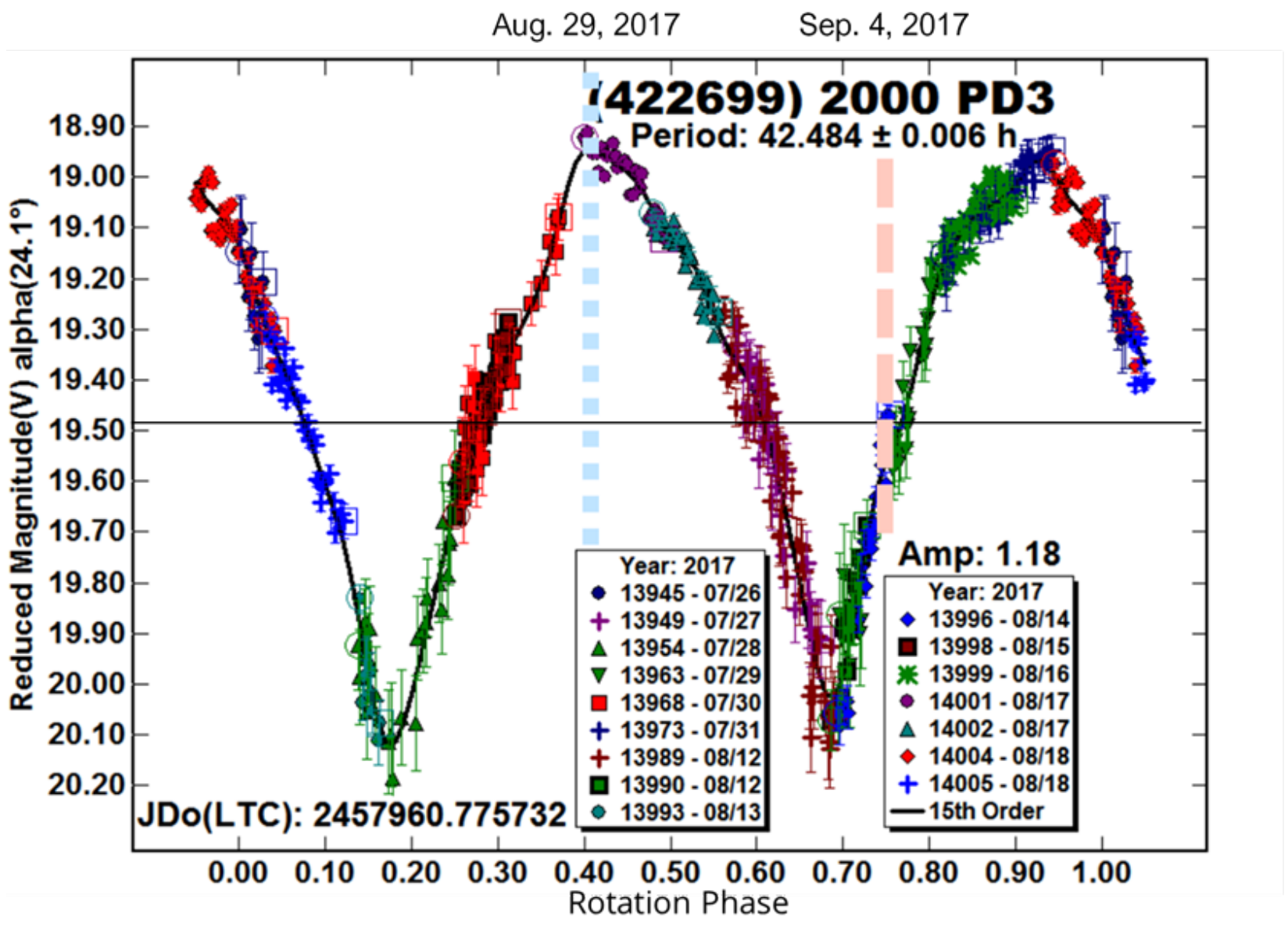

Figure 2.  Rotational phase at observing time of 2000 PD3 on August 29, 2017 (JD2457995.146) and on September 4, 2017 (JD2458001.063). The first observing run was conducted at 0.42 phase which is correspond to near the maximum of its lightcurve. The second observing run was performed at 0.76 phase around middle brightness. 0.42 phase and 0.76 phase are indicated by a blue dotted line and a pink broken line, respectively. The authority to use this plot is granted by a coauthor, Brian D. Warner.

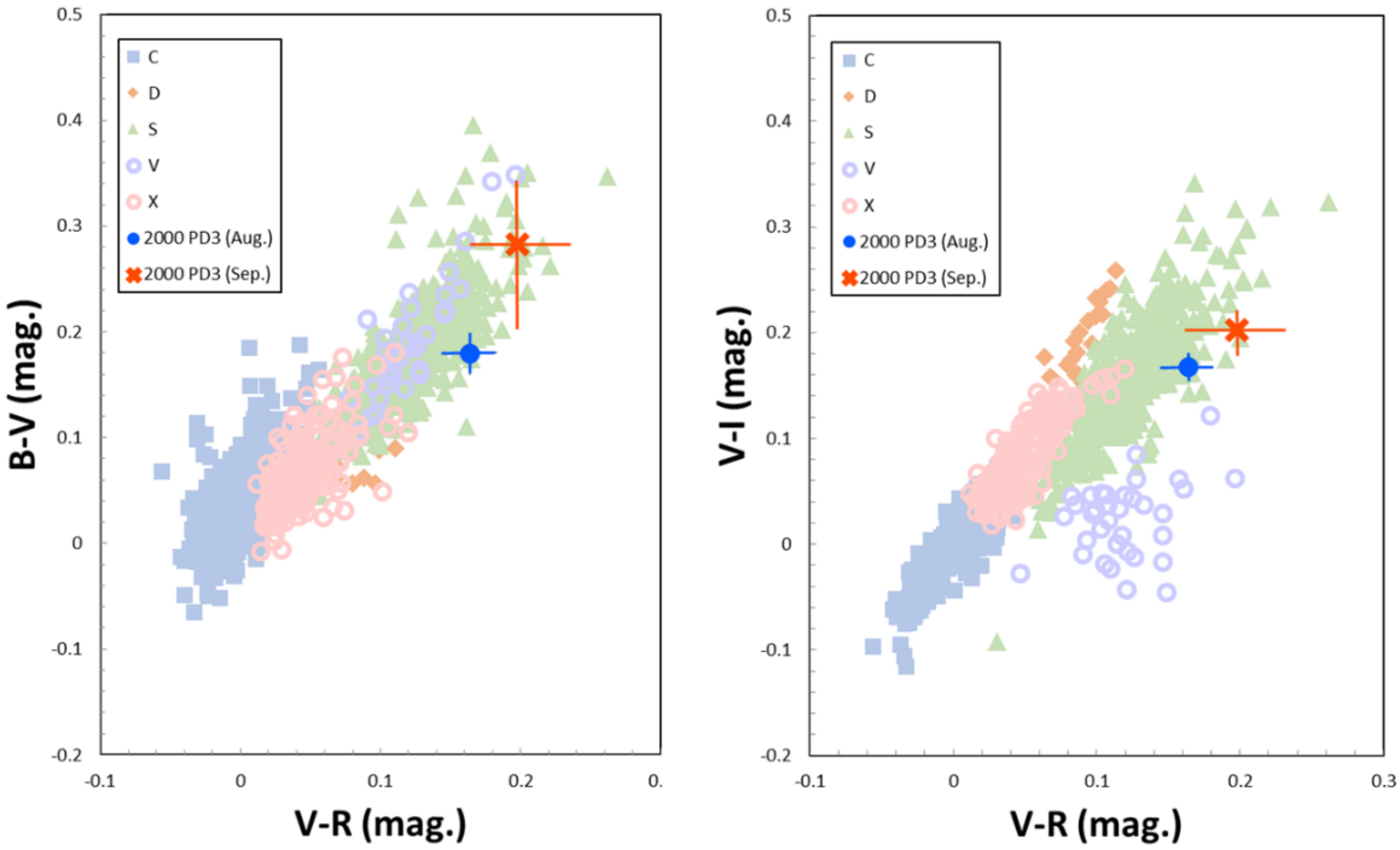

Figure 3. Color – color diagrams of each taxonomic group of asteroids and 2000 PD3. Asteroids data of five taxonomy types (C, D, S, V, X) were plotted from Bus and Binzel (2002). 2000 PD3 was marked by the filled blue circle (the first observing run on August 29, 2017) and the red cross with error bars (the second observing on September 4, 2017).

## 4. Discussion

### *4.1 2000 PD3*

From polarimetric data, 2000 PD3 is classified into S- or Q-type asteroids (an S-complex group). In addition, color photometric results support S-type taxonomy. Using the data over the near infrared from online databases (The MIT-UH-IRTF Joint Campaign for NEO Spectral Reconnaissance ) and the MIT-SMASS online tool for Bus-DeMeo taxonomy , this asteroid is classified as Sq-type (intermediate spectra between S- and Q-types). The small inversion angles may be caused by either insufficient accuracy of our data points or real physical nature of the surface grain size. Belskaya et al. (2017) presented an update plot $P_{\mathrm{min}}$ vs. $\alpha_{\mathrm{inv}}$ and showed the domains for bare rocks and lunar fines according to Dollufus et al. (1989). That indicates the low $\alpha_{\mathrm{inv}}$ tends to be large grain size. Therefore, this asteroid might be covered with large grains.

However, the color of the second observing data is different from that of the first observing data beyond their error range although both observations were carried out in the same setting. This difference can be explained by the difference of each rotational phase. Warner (2018) reported its rotation period of 2000 PD3 to be 42.484 ± 0.006 hours and lightcurve amplitude ~1.18 magnitude from visible lightcurve observation (Fig. 2). Our first run (JD2457995.146) in August was performed at 0.42 rotation phase around the maximum brightness, and our second run (JD2458001.063) in September at 0.76 rotation phase near the middle point of its lightcurve rotation from the phase zero JD2457960.776. The color difference between the large cross section side in August and the medium cross section side in September may indicate that an inhomogeneity of the chemical composition and/or freshness (i.e. space weathering degree) of its surface. The lightcurve amplitude of 1.18 corresponds to the axial ratio a/b ~ 2.96 as the minimum value which implies very elongated shape of 2000 PD3. A formation scenario that such relatively small and irregular shape asteroids like as most NEAs are fragments of parent asteroids can support for the results of surface heterogeneity and very elongated shape of 2000 PD3.

### *4.3 Phaethon*

The high polarization degree indicates that Phaethon is one of the lowest albedo asteroids which are corresponding to the F-type asteroids (e.g., Belskaya et al., 2017). Asteroids previously classified as F belong today mostly to the larger B or C classes (e.g., Bus and Binzel 2002).

Recently, Ito et al. (2018), Devogèle et al. (2018), Shinnaka et al. (2018) and Zheltobryukhov et al. (2018) made polarimetric observations of Phaethon in the very wide range of phase angle (Fig. 4). They also revealed the strong polarization of Phaethon. The maximum polarization degree $P_{\max}$ of the polarization curves in 2017 by Devogèle et al. (2018) and Shinnaka et al. (2018) were about 10 % smaller than the $P_{\max}$ in 2016 by Ito et al (2018). Small but significant discrepancies of $P_{\max}$ between 2016 and 2017 were found although the difference of $P_{\mathrm{r}}$ at phase angles lower than 60 degree is not found in Zheltobryukhov et al. (2018). However, Zheltobryukhov's results do not show the distinct difference from Ito' s results due to the large error (± 2%). Our observational results in 2017 are also inconsistent with those in 2016 by Ito et al. (2018) with an accuracy of these measurements (i.e., 0.1%,) (Fig. 4). Generally, it is possible that two or more observations on the same target in the same manner under the same condition yield different results. This can be sometimes caused by incomplete error estimate of each (or all) of the observations, especially systematic errors inherent to each instrument. In case between this study and Ito et al. (2018), because we employed the same instruments and analysis method, we conjecture that the difference of polarimetric properties between 2016 and 2017 are real. The significant difference of Phaethon's scattering properties between 2016 and 2017 may be caused by the difference of apparent angle of the object from the Earth as long as sudden alteration of the surface circumstance can be ruled out in one year.

Hanus et al. (2016) have shown ecliptic coordinates of the preferred pole orientation of (319°, -39°) with a 5° uncertainty. Phaethon exhibited the north-pole region in 2017 and the edge-on region in 2016 (see Shinnaka et al., 2018) The surface inhomogeneity of effective particle size and/or geometric albedo can give rise to a diversity of surface light-scattering property.

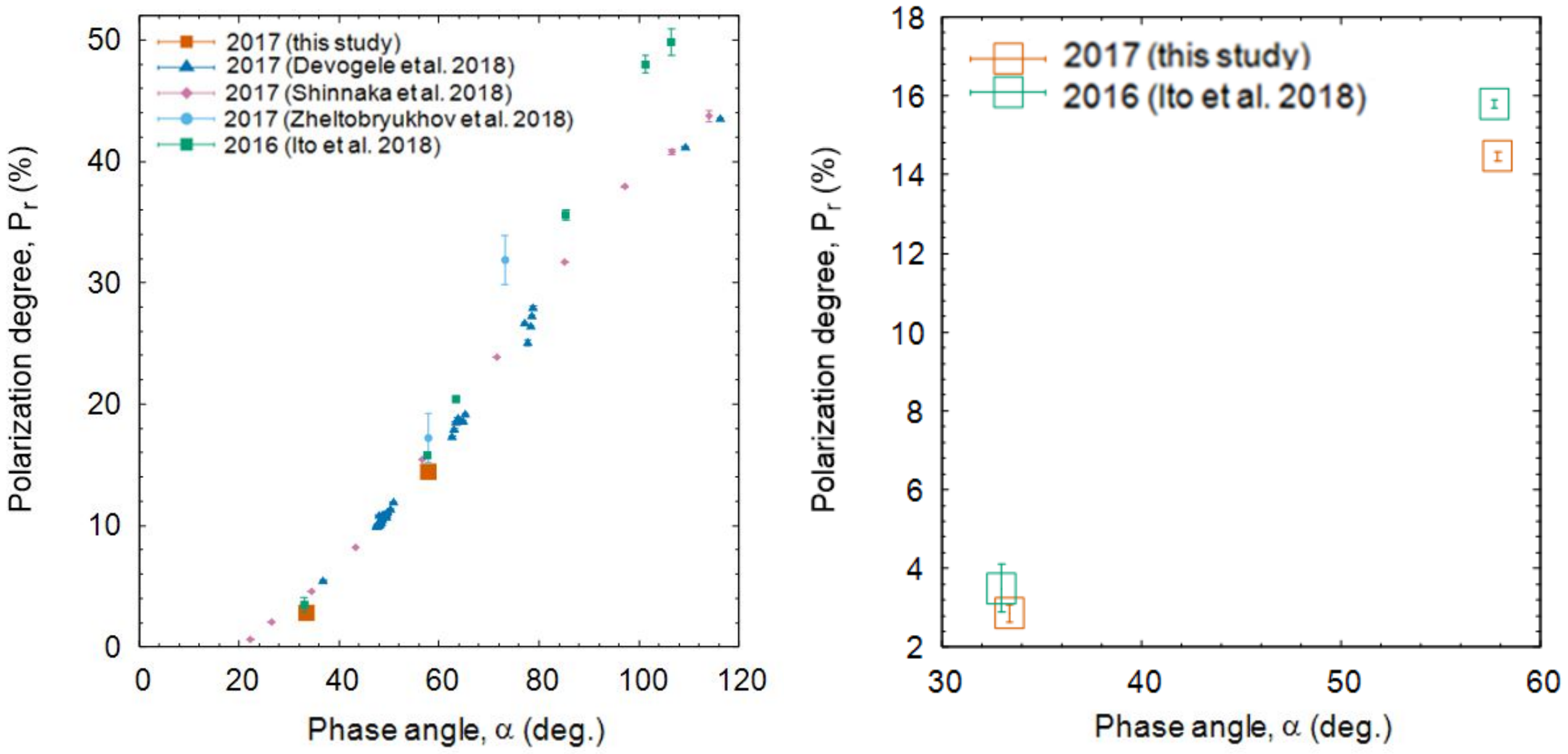

Figure 4. Polarization degree $P_r$ of Phaethon in 2016 and 2017. These data were obtained in the $R_c$-band. The orange square symbols show $P_r$ of this study observed in 2017 with error bars. The blue triangle, the pink diamond and the circle symbols display $P_r$ observed in 2017 (Devogèle et al., 2018; Shinnaka et al., 2018; Zheltobryukhov et al., 2018) and the green square symbols denote $P_r$ observed in 2016 (Ito et al., 2018). The right panel is a comparison of Phaethon's polarization degrees $P_r$ in 2017 (this study) and in 2016 (Ito et al., 2018). These data were obtained in the same method (e.g. instrument and analysis). The orange square symbols show $P_r$ of this study observed in 2017 with error bars and the green square symbols does $P_r$ observed in 2016.

## 5. Summary

Polarimetric data of near-Earth asteroids 2000 PD3 and Phaethon were obtained at very large phase angles ($25.5° \leq \alpha \leq 85.9°$). From the analysis of their surface scattering properties, 2000 PD3 shows moderate albedo ($p_v = 0.22 \pm 0.26$) and an S or Q-type asteroid. The effective diameter of 2000 PD3 was calculated as $0.69 \pm 0.15$ km. BVRI color indices demonstrate S-complex classification with the color difference due to an inhomogeneity of the chemical composition and/or freshness. Polarimetric results of Phaethon show the extremely high polarization degree. We found that our polarimetric data of Phaethon in 2017 is deviated from the previous polarimetric profile in 2016. This result supports the idea that Phaethon would have a regional heterogeneity of grain size and/or albedo on the surface.

*Acknowledgements*

The authors thank the all staff members at the Nayoro Observatory for their support in the observations. We would like to express the appreciation to the colleague, Masaya Masui and Akari Kamada for their help regarding color photometric analysis.

Funding for observations at CS3 and work on the asteroid lightcurve database (Warner et al., 2009) and ALCDEF database (alcdef.org) are supported by NASA grant 80NSSC18K0851.

Part of the discussion (spectral analysis of 2000 PD3) utilized in this publication were obtained and made available by the MIT-UH-IRTF Joint Campaign for NEO Reconnaissance. The IRTF is operated by the University of Hawaii under Cooperative Agreement no. NCC 5-538 with the National Aeronautics and Space Administration, Office of Space Science, Planetary Astronomy Program. The MIT component of this work is supported by NASA grant 09-NEOO009-0001, and by the National Science Foundation under Grants Nos. 0506716 and 0907766

Taxonomic type results presented in this work were determined, in part of the discussion (spectral analysis of 2000 PD3), using a Bus-DeMeo Taxonomy Classification Web tool by Stephen M. Slivan, developed at MIT with the support of National Science Foundation grant 0506716 and NASA grant NAG5-12355.

*Appendix*

The observed ordinary and extraordinary fluxes at the halfwave plate angle Ψ in degrees, $I_o$ (Ψ) and, $I_e$ (Ψ), were used to derive

$$q'_{pol} = \left(\frac{R_q - 1}{R_q + 1}\right) / P_{eff} \tag{12}$$

and

$$u'_{pol} = \left(\frac{R_u - 1}{R_u + 1}\right) / P_{eff} \tag{13}$$

where $R_q$ and $R_q$ are obtained from each observation using the following equations:

$$R_q = \sqrt{\frac{I_e(0)/I_o(0)}{I_e(45)/I_o(45)}} \tag{14}$$

and

$$R_u = \sqrt{\frac{I_e(22.5)/I_o(22.5)}{I_e(67.5)/I_o(67.5)}} \tag{15}$$

where $P_{eff}$ is a polarization efficiency, which was examined by taking dome flat images through a pinhole and a Polaroid-like linear polarizer, which produces artificial stars with P = 99.97 ± 0.02% (V) and 99.98 ± 0.01% ($R_c$). $P_{eff}$ was measured approximately two months prior to our observation and was determined to be $P_{eff}$ = 99.32 ± 0.03 % in the V-band and 99.42 ± 0.03 % in the $R_c$-band. The instrumental polarization of Pirka/MSI is known to depend on the instrument angle of rotation and can be corrected with the following equation:

$$\begin{pmatrix} q''_{pol} \\ u''_{pol} \end{pmatrix} = \begin{pmatrix} q'_{pol} \\ u'_{pol} \end{pmatrix} - \begin{pmatrix} \cos 2\theta_{rot1} & -\sin 2\theta_{rot1} \\ \sin 2\theta_{rot2} & \cos 2\theta_{rot2} \end{pmatrix} \begin{pmatrix} q_{inst} \\ u_{inst} \end{pmatrix} \tag{16}$$

where $\theta_{rot1}$ denotes the average instrument rotator angle during the exposures with Ψ = 0° and 45°, while $\theta_{rot2}$ denotes the average angle with Ψ = 22°.5 and 67°.5. $q_{inst}$ and $u_{inst}$ are two components of the Stokes parameters for the instrumental polarization and were determined to be $q_{inst}$ = 0.963 ± 0.029% in the V-band and 0.703 ± 0.033% in the $R_c$-band and $u_{inst}$= 0.453 ± 0.043% in the V-band and 0.337 ± 0.020% in the $R_c$-band, respectively, by observing the unpolarized stars HD 212311 and BD+32 3739 (see Table 3, on page 1566, Schmidt et al. 1992). The instrument position angle in celestial coordinates was determined by measuring the polarization position angles of strongly polarized stars for which position angles are reported in Schmidt et al. (1992). The instrument position angle can be corrected using the following equations:

$$\begin{pmatrix} q'''_{pol} \\ u'''_{pol} \end{pmatrix} = \begin{pmatrix} \cos 2\theta'_{off} & \sin 2\theta'_{off} \\ -\sin 2\theta'_{off} & \cos 2\theta'_{off} \end{pmatrix} \begin{pmatrix} q''_{pol} \\ u''_{pol} \end{pmatrix} \tag{17}$$

and

$$\theta'_{off} = \theta_{off} - \theta_{ref} \tag{18}$$

where $\theta_{ref}$ is a given parameter for specifying the position angle of the instrument and determined to be $\theta_{ref}$ = -0.52. Through an observation of strongly polarized stars (BD+59 389, HD19820 & HD25443 ) in 2016 Oct, we derived $\theta_{off}$ = 3.93 ± 0°.10 in the V-band and 3°.94 ± 0°.31 in the $R_c$-band.